\documentclass[onecolumn%
 reprint,
 aps,
]{revtex4-2}

\usepackage{graphicx}% Include figure files
\usepackage{dcolumn}% Align table columns on decimal point
\usepackage{bm}% bold math
\begin{document}

%\preprint{APS/123-QED}

\title{Quantum secure data transfer with pulse shape encoded optical qubits}% Force line breaks with \\

\author{Rui-Xia Wang}

 \email{wangrx.2009@tsinghua.org.cn}
\affiliation{School of Natural Sciences, University of California, Merced, California 95343, USA}%

%\affiliation{
%\affiliation{
%\affiliation{%


\begin{abstract}

Quantum secure data transfer is an important topic for quantum cyber security. We propose a scheme to realize quantum secure data transfer in the basis of quantum secure direct communication (QSDC). In this proposal, the transmitted data is encoded in the pulse shape of a single optical qubit, which is emitted from a trapped atom owned by the sender and received by the receiver with another trapped atom. The encoding process can be implemented with high fidelity by controlling the time-dependent driving pulse on the trapped atom to manipulate the Rabi frequency in accordance with the target pulse shape of the emitted photons. In the receiving process, we prove that, the single photon can be absorbed with arbitrary probability by selecting appropriate driving pulse. We also show that, based on the QSDC protocol, the data transfer process is immune to the individual attacks.

%\begin{description}
%\end{description}
\end{abstract}

\maketitle

%\tableofcontents

%\section{\label{sec:level1}First-level heading:\protect\\ The line
%break was forced \lowercase{via} \textbackslash\textbackslash}

%\section{\label{sec:level1}introduction }

The development of large quantum computers will have dire consequences on cyber security, the security of data transfer becomes a pressing issue, however, quantum technologies will also have positive impact on it and the protection of data transfer is investigated by many experts and researchers\cite{mosca2018cybersecurity,wallden2019cyber,abd2020providing,chowdhury2021physical}. Quantum communication provides unconditional security in principle for exchanging information over public channel based on quantum mechanics. It is impossible to eavesdrop without disturbing the quantum transmission. There are many protocols for quantum communication, such as the quantum key distribution\cite{2014Quantum,Ekert1991Quantum,2001Unconditional}, quantum secret sharing\cite{1999Quantum}, quantum telepotation\cite{1993Teleporting,Bouwmeester1997Experimental}, and quantum secure direct communication (QSDC)\cite{2002Theoretically,2003Two,2004Secure,2005Quantum,2008Deterministic,2009Confidential,2017Quantum,Peng2018Measurement,2019Implementation}. Among these protocols, QSDC can communicate information directly without key distribution, which can further enhance the security of the quantum communication by eliminating the security loopholes with key storage or ciphertext attacks\cite{Peng2018Measurement,2019Implementation}. QSDC was proposed in 2000\cite{2002Theoretically}, and it has been realized in the experiment by encoding the information to the photon spin states\cite{2019Implementation,2017Experimental}. The security analysis for QSDC has been done by combining the quantum mechanics with Wyner's wiretap channel theory\cite{2019Security}. The security of data transfer can also be realized by encoding the transmitted data in the quantum state based on the protocol of QSDC.

In the quantum state transfer, the optical photons can serve as ideal flying qubits at room temperature for they having very weak coupling with the environment at room temperature\cite{2001Quantum}. Photons can carry and transmit quantum information between different quantum devices. Atoms are well suited for storing qubits in long-lived internal states, the stored quantum information can be processed locally using quantum gates. Single atoms trapped in high-Q cavities can compose a quantum device which are capable of storing, sending and receiving quantum information which is encoded in an optical qubit\cite{2008The,2010Colloquium,2012An}. Based on the time-reversal symmetry of the trapped atom system, the time-dependent driving pulse can be obtained for receiving a single photon with high fidelity\cite{1996Quantum,2003Cavity,2006Universal,Fleischhauer2000How,2007Optimal,2012Single}. The exchange of information between different quantum devices can be accomplished with the flying qubits of single photons transmitted via optical fibers.

In the previous experimental quantum communication protocols, the information is encoded in various degrees of freedom such as the polarization of the photons\cite{2016Experimental}, the phase of the particle\cite{2018Secure,2019Experimental} and time-bin states\cite{2017Provably,2019Genuine}. Here we propose a scheme to realize quantum secure data transfer in the basis of QSDC protocol with trapped atoms. The single photons emitted from the trapped atoms are in time-bin states. Quantum information is encoded in the relative phase of the time bins of the single photons and there are different pulse shapes for encoding the same quantum information, which are used for the eavesdropping checking. The single photon pulse shape can be controlled by the time-dependent driving pulse acting on the trapped atoms. In the absorbing process, controllable absorbing probability can be obtained with an appropriate optimal time-dependent driving pulse based on the time-reversal symmetry of the system.

\begin{figure}[h]
\centering{\includegraphics[width=80mm]{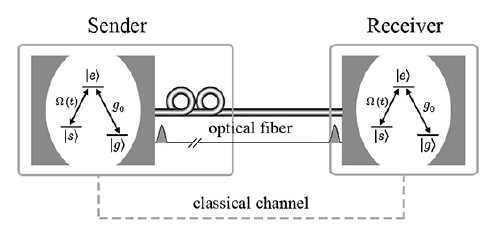}}
\caption{Schematic diagram for quantum secure data transfer. The devices for both the sender and the receiver are composed of n atom-cavity systems respectively, and the systems are one by one correspondence. The sender transmits quantum information encoded in the single photons over the optical fibers. The length of the fiber at the sender side is much longer than the receiver. The two sides can also communicate with a classical channel for eavesdropping checking.}
\label{ab}
\end{figure}

Our model contains one atom with three energy levels which is trapped
in an optical cavity. There are two ground states $|g\rangle$ and $|s\rangle$ and an excited
state $|e\rangle$ for the trapped atom. The transition $|e\rangle$ to $|g\rangle$ is coupled resonantly to the cavity mode $a$ with a strong coupling rate $g_0$\cite{2010Single,2011The,2011Single,haroche2013nobel}. The atom-cavity systems compose a quantum device for storing and sending quantum information. First, we consider a closed system, the simplified Hamiltonian for each atom-cavity system
is $H_{I}=\Omega\left(t\right)|e\rangle\langle s|+g_{0}|e\rangle\langle g|\hat{a}+H.c.$ $\left(\hbar=1\right)$, where $\Omega\left(t\right)$ is the Rabi oscillation frequency of the transition $|e\rangle$ to $|s\rangle$, which is proportional to the strength of the time-dependent driving pulse to the transition $|e\rangle$ to $|s\rangle$. So that, $\Omega(t)$ is tunable by varying the driving field. $\hat{a}$ is the destruction operator for cavity mode $a$. This
Hamiltonian has three eigenstates, one is a dark state $|D\rangle$ with the eigenvelue of $\lambda_{D}=0$, the others
are bright states $|B_{+}\rangle$ and $|B_{-}\rangle$. Define that $\cos\theta\left(t\right)=\frac{\Omega\left(t\right)}{\sqrt{g_{0}^{2}+\Omega\left(t\right)^{2}}}$
and $\sin\theta\left(t\right)=\frac{g_{0}}{\sqrt{g_{0}^{2}+\Omega\left(t\right)^{2}}}$
, in the bases of $|s\rangle|0\rangle$, $|e\rangle|0\rangle$ and $|g\rangle|1\rangle$,
where $\left|0\right\rangle $ and $\left|1\right\rangle $ represent
the zero- and the one-photon state of the cavity mode $a$ respectively, the dark state is
\begin{equation}
|D\rangle=-\cos\theta(t)|g\rangle|1\rangle+\sin\theta(t)|s\rangle|0\rangle.
\label{ds}
\end{equation}

Assume that, the wave function of the closed system is $\psi(t)=c_1(t)|s\rangle|0\rangle+c_2(t)|e\rangle|0\rangle+c_3(t)|g\rangle|1\rangle$. Initially, if the system is in state $|s\rangle|0\rangle$ or $|g\rangle|1\rangle$, in the adiabatic process, the system will evolve in the dark state, which means that $c_2(t)=0$ in the whole process. In the bases of $|D\rangle$, $|B_+\rangle$ and $|B_-\rangle$, the wave function is $\psi(t)=c_D(t)|D\rangle+c_{B+}(t)|B_+\rangle+c_{B-}(t)|B_-\rangle$, initially, if the system is in state $|D\rangle$, under the adiabatic approximation, there will be $c_{B+}(t)=0$ and $c_{B-}(t)=0$.

Consider the coupling between the cavity mode $a$ and the cavity output or input, $\kappa$ is the cavity decay rate,  the Heisenberg-Langevin function for this open system is $\dot{\hat{a}}\left(t\right)=-i\left[\hat{a}\left(t\right),H_{I}\right]-\frac{\kappa}{2}\hat{a}\left(t\right)+\sqrt{\kappa}\hat{a}_{in}\left(t\right)$.

The input and output relation is $\hat{a}_{in}\left(t\right)+\hat{a}_{out}\left(t\right)=\sqrt{\kappa}\hat{a}\left(t\right)$. In the output process, we assume that $\langle\hat{a}_{in}\left(t\right)\rangle=0$, we can get a new function $\left\langle \hat{a}_{out}\left(t\right)\right\rangle =\sqrt{\kappa}\left\langle \hat{a}\left(t\right)\right\rangle$. And the conditional Hamiltonian of the system for the output process is $H_{c}=-i\frac{\kappa}{2}\hat{a}^{\dagger}\hat{a}+\left[\Omega\left(t\right)|e\rangle\langle s|+g_{0}|e\rangle\langle g|\hat{a}+H.c.\right]$.

In the bases of $|D\rangle$, $|B_+\rangle$ and $|B_-\rangle$, substituting the wave function to the schr$\ddot{o}$dinger equation $i\frac{\partial}{\partial t}|\psi(t)\rangle=H_c|\psi(t)\rangle$, under the adiabatic approximation, the simplified function for $c_D$ is (see Appendix A and reference \cite{2003Cavity} for more details) $\dot{c}_{D}(t)=-\frac{\kappa}{2}\cos^{2}\theta(t) c_{D}(t)$. The straightforward solution for this function is $c_{D}(t)=c_{D}\left(0\right)\exp\left[-\frac{\kappa}{2}\int_{0}^{t}\cos^{2}\theta\left(\tau\right)d\tau\right]$. The relationship between $c_3(t)$ and $c_D(t)$ can be obtained from equation \ref{ds} as $c_{3}(t)=-\cos\theta(t)c_{D}(t)$. In this open system, $c_3(t)=\langle\hat{a}(t)\rangle$, then we can get the output pulse as $\langle \hat{a}_{out}(t)\rangle =-\sqrt{\kappa}\cos\theta(t) c_{D}\left(0\right)\exp\left[-\frac{\kappa}{2}\int_{0}^{t}\cos^{2}\theta\left(\tau\right)d\tau\right]$, where $\theta(t)$ is a function of $\Omega(t)$. Using the optimization method, for every target output single photon pulse shape $\left\langle \hat{a}_{out}\left(t\right)\right\rangle $, we can find a corresponding $\Omega\left(t\right)/g_{0}$.

\begin{figure}[t]
\centering{\includegraphics[width=80mm]{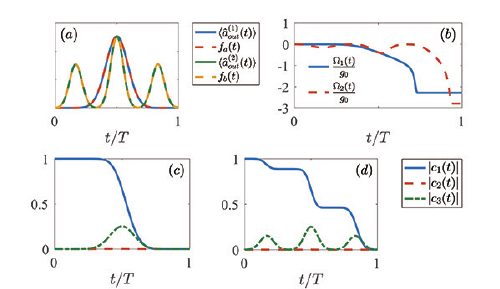}}
\caption{The optimization and numerical simulation results for the single photon output process. (a) $f_a(t)$ and $f_b(t)$ ($t\in [0,T]$) are the first and second target functions for the single photon pulse. $f_a(t)$ which satisfies a Gaussian function and there is $\int_0^{T}|f_a(t)|^2 dt=1$. $\langle\hat{a}_{out}^{(1)}(t)\rangle$ is the numerical simulation results of the output pulse with $f_a(t)$ as the target function, which is driven by the optimal $\Omega_1(t)/g_0$ shown in (b). For $f_b(t)$, the former, middle and later parts this function in time domain satisfy three different Gaussian functions respectively. There still is $\int_0^{T}|f_b(t)|^2 dt=1$. $\langle\hat{a}_{out}^{(2)}(t)\rangle$ is the numerical simulation results of the output pulse with $f_b(t)$ as the target function, which is driven by the optimal $\Omega_2(t)/g_0$ shown in (b). In the optimization process for $\Omega_1(t)/g_0$ and $\Omega_2(t)/g_0$, we assume $\kappa=100/T$. The target functions $f_a(t)$ and $f_b(t)$ satisfy the relation $\int_0^{T}f_a(t)f_b(t)dt\approx 1/\sqrt{2}$, $f_a(t)$ and $f_b(t)$ only have real part. (c) and (d) are the numerical simulation results for the output process with $\Omega_1(t)/g_0$ and $\Omega_2(t)/g_0$ respectively, where $g_0=100\kappa$.}
\label{fig2}
\end{figure}

Apart from $\Omega(t)$, the coupling strength $g_0$ will also produce an effect on the fidelity of the whole process. In different cases of $g_{0}$, $c_{1}\left(t\right)$,
$c_{2}\left(t\right)$ and $c_{3}\left(t\right)$ can be calculated by substituting
$\Omega(t)$ and $g_0$ into the equation $\dot{\vec{v}}=-iM\vec{v}$, where $\vec{v}=[c_1(t),c_2(t),c_3(t)]^{T}$ and $M=[0,\Omega(t),0;\Omega(t),0,g_0;0,g_0,-\frac{i\kappa}{2}]$. The optimization and numerical simulation results for the output process is shown in Figure $\ref{fig2}$.

We can prove that, this system is time-reversible symmetric (see Appendix B for more details). With the strong coupling between the atom and the cavity,
the input and output process is reversible. In the absorbing process, initially, the system is in state $|g\rangle|0\rangle$. If the target input function is $f_a(t)$ ($t\in [0,T]$), we can get the time-dependent Rabi frequency in the absorbing process by doing as follow. 

First, we can assume that, the system is in state $|s\rangle|0\rangle$. And then, optimize the output process with a target output pulse function $f_a(T-t)$ and getting an optimized result for $\Omega(t)/g_0$. Finally, we can absorb an input pulse $f_a(t)$ with $\Omega(T-t)/g_0$. Assume that, following these steps, the ratio of the Rabi frequency over $g_0$, which we find in the absorbing process, is $\Omega_a(t)/g_0$.

Under the adiabatic approximation, in the absorbing process, the system will stay in the dark state. The decay rate for the dark state is $\kappa_{D}=\kappa\cos^{2}\theta$,
then we have $\dot{c_{D}}(t)=-\frac{\kappa}{2}\cos^{2}\theta(t) c_{D}(t)+\sqrt{\kappa}\cos\theta(t) \langle\hat{a}_{in}(t)\rangle$. Solving this function gives the solution $c_{D}(t)=c_{D}(0)\exp\left[-\frac{\kappa}{2}\int_{0}^{t}\cos^{2}\theta\left(t^{\prime}\right)dt^{\prime}\right]\nonumber+\int_{0}^{t}J(t,t^{\prime})\langle\hat{a}_{in}\left(t^{\prime}\right)\rangle dt^{\prime}$, where $J(t,t^{\prime})=\sqrt{\kappa}\cos\theta\left(t^{\prime}\right)\exp\left[\frac{\kappa}{2}\int_{t}^{t^{\prime}}\cos^{2}\theta\left(t^{\prime\prime}\right)dt^{\prime\prime}\right]$ and $\langle\hat{a}_{in}\left(t^{\prime}\right)\rangle=f_a(t^{\prime})$.

In the absorbing process, given the constant parameters $g_0$ and $\kappa$, an input pulse $f_a(t)$ can be absorbed by the optimized time-dependent Rabi frequency $\Omega_a(t)$. Ideally, there is $c_D^{aa}(0)=0$ and $c_D^{aa}(T)=1$, where the first $a$ in the superscript of $c_D^{aa}$ is the same as the subscript of $f_a(t)$ and the second $a$ is the same as the subscript of $\Omega_a(t)$. Then we can get $\int_{0}^{T}J(T,t^{\prime})\langle\hat{a}_{in}\left(t^{\prime}\right)\rangle dt^{\prime}=1$, where $J(T,t^{\prime})=\sqrt{\kappa}\cos\theta_a\left(t^{\prime}\right)\exp\left[\frac{\kappa}{2}\int_{T}^{t^{\prime}}\cos^{2}\theta_a\left(t^{\prime\prime}\right)dt^{\prime\prime}\right]$ and $\tan{\theta_a(t)}=g_0/\Omega_a(t)$.

We can assume that, the input pulse function is normalized, there is $\int_{0}^{T}|\langle\hat{a}_{in}\left(t^{\prime}\right)\rangle|^{2}dt^{\prime}=1$,
we can prove that, at any time $t^{\prime}$, there is $J(T,t^{\prime})=\langle\hat{a}_{in}^{\star}\left(t^{\prime}\right)\rangle=f_a^{\star}(t^{\prime})$,
where $a^{\star}(t^{\prime})$ ($f_a^{\star}(t^{\prime})$) is conjugate value of $a(t^{\prime})$ ($f_a(t^{\prime})$) (see Appendix C for more details). If the input pulse function is not normalized and $\int_{0}^{T}|\langle\hat{a}_{in}\left(t^{\prime}\right)\rangle|^{2}dt^{\prime}=x$ ($x\in R_{+}$), there is $J(T,t^{\prime})=\langle\hat{a}_{in}^{\star}\left(t^{\prime}\right)\rangle/\sqrt{x}=f_a^{\star}(t^{\prime})/\sqrt{x}$. So that, under the strong coupling condition, with the constant value of $g_0$ and $\kappa$, in the absorbing process, if the normalized input pulse is changed from $f_a(t)$ to $f_b(t)$ ($t\in [0,T]$), but the Rabi frequency is still $\Omega_a(t)$, the relationship between the final value of $c_{D}^{ba}$
and the overlap of the two input pulse functions can be calculated as follows.

\begin{figure}[h]
\centering{\includegraphics[width=80mm]{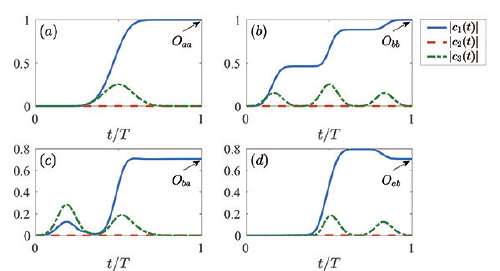}}
\caption{Numerical simulation results for the absorbing process. (a) is the process of absorbing a photon with pulse shape $f_a(t)$ and the Rabi frequency $\Omega_a(t)$. At time $t=T$, $c_3(T)\approx1$. (b) is the process of absorbing a photon with pulse shape $f_b(t)$ with the Rabi frequency $\Omega_b(t)$. At time $t=T$, $c_3(T)\approx1$. (c) is the process of absorbing a photon with pulse shape $f_b(t)$ with the Rabi frequency $\Omega_a(t)$. At time $t=T$, $c_3(T)\approx1/\sqrt{2}$. (d) is the process of absorbing a photon with pulse shape $f_a(t)$ with the Rabi frequency $\Omega_b(t)$. At time $t=T$, $c_3(T)\approx1/\sqrt{2}$.}
\label{fig3}
\end{figure}

Define the overlap between two normalized wave functions $f_a(t)$ and $f_b(t)$ ($t\in[0,T]$) as $O_{ba}(0,T)=\int_{0}^{T}f_a^{\star}\left(t\right)f_b\left(t\right)dt$. In the absorbing process, there is $c_{D}^{ba}\left(T\right)= \int_{0}^{T}f_a^{\star}(t^{\prime})f_b(t^{\prime})dt^{\prime}=O_{ba}$, $\theta_i(T)$ ($i=a,b...$) is equal to the value of $\theta(t)$ at $t=0$ in the photon-emitting process. So that, there is $\theta_i(T)=1$ and $c_3^{ij}=c_D^{ij}$ ($i,j=a,b...$). Figure \ref{fig3} shows the numerical simulation results of absorbing process. Because $f_i(t)$ ($i=a,b$) is real, and there is $\int_0^{T}f_a(t)f_b(t)dt\approx1/\sqrt{2}$, we can get $O_{ab}=O_{ba}\approx1/\sqrt{2}$. Figure \ref{fig3}(c) and (d) show that, at time $t=T$, $c_3(T)\approx1/\sqrt{2}$, which means that, in this case, there is $\frac{1}{2}$ probability to absorb a single photon into the trapped atom.

\begin{figure}[h]
\centering{\includegraphics[width=80mm]{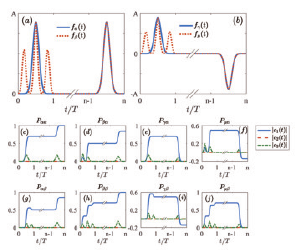}}
\caption{Numerical simulation results for the quantum secure data transfer process. (a) and (b) are four types of single photon pulse shapes used in the quantum information transfer. (c) to (j) show the process of absorbing a single photon by the receiver. $P_{ij}$ is the process of absorbing a single photon with pulse shape $f_i(t)$ with a time-dependent Rabi frequency $\Omega_j(t)$. For the processes $P_{\alpha\alpha}$, $P_{\gamma\alpha}$, $P_{\beta\beta}$ and $P_{\mu\alpha}$, during time $t\in(T,(n-1)T)$, there is $50\%$ probability to absorb a single photon. While for the processes $P_{\beta\alpha}$, $P_{\mu\alpha}$, $P_{\alpha\beta}$ and $P_{\gamma\alpha}$, during time $t\in(T,(n-1)T)$, there is only $25\%$ probability to absorb a single photon. For the processes $P_{\alpha\alpha}$ and $P_{\beta\beta}$, at time $t=nT$, there is unit probability to absorb a single photon, and the receiver can receive the information of $1$. For the processes $P_{\gamma\alpha}$ and $P_{\mu\beta}$, at time $t=nT$, there is no probability to absorb a single photon, and the receiver can receive the information of $0$.}
\label{fig4}
\end{figure}

This system can be used to realize quantum secure state transfer.
The quantum devices for both sender and receiver are composed of n trapped atoms. Here, we assume that, the fiber used to transmit the single photon is lossless. There are four different pulse shapes $f_j(t)$ ($j=\alpha,\beta,\gamma,\mu$ and $t\in[0,nT]$ and $n>1$) for a single photon used in the quantum data transfer, which are shown in figure \ref{fig4}(a) and (b). For simplicity, we assume $f_j(t)$ is real. $f_{\alpha}$ and $f_{\gamma}$ are used to transmit the information $1$, the others are used as the information $0$. The four types of pulse satisfy $\int_0^{nT}|f_j(t)|^2dt=1$, $\int_0^{T}|f_j(t)|^2dt=1/2$ and $\int_{(n-1)T}^{T}|f_j(t)|^2dt=1/2$. During the time $t\in(T,(n-1)T)$, there is $f_j(t)\approx0$ and during time $t\in[0,T]$, there are $\int_0^{T}f_\alpha(t)f_{\beta}(t)dt=\frac{1}{2\sqrt{2}}$ and $\int_0^{T}f_\gamma(t)f_{\mu}(t)dt=\frac{1}{2\sqrt{2}}$. The time-dependent Rabi frequency $\Omega_j(t)$ is optimized to absorb the input pulse $f_j(t)$ with the unit probability. The wave function for $f_j(t)$ can be written as $|\Phi\rangle=\frac{1}{\sqrt{2}}|x_j\rangle+\frac{1}{\sqrt{2}}e^{i\phi}|y_j\rangle$, where $|x\rangle$ represents the state during $t\in[0,T]$ and $|y\rangle$ the state in $t\in[(n-1)T,nT]$. For $j=\alpha,\gamma$, $\phi=0$ and for $j=\beta,\mu$, $\phi=\frac{\pi}{2}$. And for $j=\alpha,\beta$, there is $\langle x_{\alpha}|x_{\beta}\rangle=1/\sqrt{2}$ and for $j=\gamma,\mu$, there is $\langle x_{\gamma}|x_{\mu}\rangle=1/\sqrt{2}$. The sender and receiver are simplified as S and R, respectively. The processes of quantum data transfer are as follows.

Step 1. The atom-cavity systems belonging to S are initialized to state $|s\rangle|0\rangle$ and which for R are initialized to state $|g\rangle|0\rangle$. The S and R's atom-cavity systems are one-to-one correspondence.

Step 2. If the data sending from S to R is $1$ $(0)$, we can apply an optimized driving pulse on the atom of S to produce a single photon with pulse shape $f_{\alpha}(t)$ ($f_{\gamma}(t)$) or $f_{\beta}(t)$ ($f_{\mu}(t)$), and transmitted to R. When the front end of the single photon pulse arrives at R, the time is marked as $t=0$. At this time, S just finishes the emitting, the single photon is totally in the optical fiber with certain pulse shape.

Step 3. The single photon can be received by applying a time-dependent driving pulse to produce a corresponding Rabi frequency, which is chosen randomly only from $\Omega_{\alpha}(t)$ and $\Omega_{\beta}(t)$ by R. If S send $f_{\alpha}(t)$ ($f_{\beta}(t)$) or $f_{\gamma}(t)$ ($f_{\mu}(t)$) to R, R chooses $\Omega_{\alpha}(t)$ ($\Omega_{\beta}$) to receive the photon, then the choice is right, if not, it is wrong.

Step 4. After $t=T$, R chooses $m$ $(m<n)$ atoms randomly to detect the states of these atoms, and will get $0$ (if the atom is in state $|g\rangle$) or $1$ (if the atom is in state $|s\rangle$).

Step 5. Eavesdropping checking: R and S communicate with classical channel for which time-dependent Rabi frequency R chooses for all the trapped atoms and which atoms are chosen to measure and the measurement results. For the $m$ atoms, if R chooses the right driving pulse, the result is marked as type $r$, if R didn't choose the right driving pulse, the result is marked as type $nr$. In type $r$, if the probability for R getting $1$ is $50\%$, and in type $nr$, the probability for R getting $1$ is $25\%$, the quantum communication is safe, If not, S should stop all the sending processes immediately.

If the quantum state transfer is safe, S will do nothing with the processes for the atoms which R chooses the right driving pulse and the later part of these single photon pulses will be continued to send to R, the number of which is marked as $q$. For the processes which R chooses wrong driving pulses in the $n-m$ atoms, S stops the sending processes and tell R which ones they are. When this checking process is finished, we mark the time as $t_r$. There are redundant optical fiber in S side, which should make sure that, the transmission time of the photon in S side is longer than $t_r+T$, so that, at time $t=t_r$, the later part of the single photons are still under control of S, and S can choose to detect the later part of the single photons to stop sending, there isn't any information sent out from S before time $t_r$. If the sending and receiving are bidirectional, there should also be redundant length of optical fiber in R side to make sure the later part of the single photons are still under control at the end of the eavesdropping checking process, it is possible for R to transfer quantum states to S securely.

Step 6. After R receives all $q$ photons with right driving pulses and detects the states of the $q$ trapped atoms, he can get the transferred information.

If there is eavesdropping after the eavesdropping checking, the state of the eavesdropper's trapped atoms is random and he will get nothing.

In this proposal, we encoded the quantum information in the pulse shape and the phase of single photons to realize the quantum secure data transfer in the basis of QSDC protocol. In the emitting process, the optimal time-dependent Rabi frequency to produce a single photon with arbitrary pulse shape can be realized with high fidelity. In the absorbing process, the relationship between the time-dependent Rabi frequency and the absorbing probability for a certain input photon pulse was calculated. With this method, a single photon can be absorbed with arbitrary probability by using the corresponding time-dependent Rabi frequency. We prove that, the quantum secure data transfer can be implemented with the atom-cavity systems.

\appendix

\section{Emitting process}

In the basis of $|D\rangle$, $|B_{+}\rangle$ and $|B_{-}\rangle$,
the conditional Hamiltonian can be written as

\begin{eqnarray}
H_{c}=\left[\begin{array}{ccc}
-i\frac{\kappa}{2}\cos^{2}\theta & \left(\frac{i\kappa}{2\sqrt{2}}+\frac{1}{\sqrt{2}}\right)\cos\theta\sin\theta & -\left(\frac{i\kappa}{2\sqrt{2}}+\frac{1}{\sqrt{2}}\right)\cos\theta\sin\theta\\
\left(\frac{i\kappa}{2\sqrt{2}}+\frac{1}{\sqrt{2}}\right)\cos\theta\sin\theta & -i\frac{\kappa}{4}\sin^{2}\theta+\sqrt{g_{0}^{2}+\Omega\left(t\right)^{2}} & -\frac{i\kappa}{4}\sin^{2}\theta-\frac{1}{2}\cos^{2}\theta+\frac{1}{2}\\
-\left(\frac{i\kappa}{2\sqrt{2}}+\frac{1}{\sqrt{2}}\right)\cos\theta\sin\theta & -\frac{i\kappa}{4}\sin^{2}\theta-\frac{1}{2}\cos^{2}\theta+\frac{1}{2} & -i\frac{\kappa}{4}\sin^{2}\theta-\sqrt{g_{0}^{2}+\Omega\left(t\right)^{2}}
\end{array}\right]
\end{eqnarray}

The wave function of the system can be written as

\begin{equation}
|\psi\left(t\right)\rangle=c_{D}\left(t\right)|D\left(t\right)\rangle+c_{+}\left(t\right)|B_{+}\left(t\right)\rangle+c_{-}\left(t\right)|B_{-}\left(t\right)\rangle
\end{equation}

We know that

\begin{eqnarray*}
|\dot{D}\rangle & = & \frac{\dot{\theta}}{\sqrt{2}}\left(|B_{+}\rangle-|B_{-}\rangle\right)\\
|\dot{B}_{+}\rangle & = & -\frac{\dot{\theta}}{\sqrt{2}}|D\rangle\\
|\dot{B}_{-}\rangle & = & \frac{\dot{\theta}}{\sqrt{2}}|D\rangle
\end{eqnarray*}

Substituting the wave function to the schr$\ddot{o}$dinger equation,
we can get

\begin{eqnarray*}
i\left(\dot{c}_{D}-\frac{\dot{\theta}}{\sqrt{2}}c_{+}+\frac{\dot{\theta}}{\sqrt{2}}c_{-}\right) & = & -i\frac{\kappa}{2}\cos^{2}\theta c_{D}+\left(\frac{i\kappa}{2\sqrt{2}}+\frac{1}{\sqrt{2}}\right)\cos\theta\sin\theta c_{+}-\left(\frac{i\kappa}{2\sqrt{2}}+\frac{1}{\sqrt{2}}\right)\cos\theta\sin\theta c_{-}\\
i\left(c_{D}\frac{\dot{\theta}}{\sqrt{2}}+\dot{c}_{+}\right) & = & \left[-i\frac{\kappa}{4}\sin^{2}\theta+\sqrt{g_{0}^{2}+\Omega\left(t\right)^{2}}\right]c_{+}+\left(\frac{i\kappa}{2\sqrt{2}}+\frac{1}{\sqrt{2}}\right)\cos\theta\sin\theta c_{D}\\
 & + & \left(-\frac{i\kappa}{4}\sin^{2}\theta-\frac{1}{2}\cos^{2}\theta+\frac{1}{2}\right)c_{-}\\
i\left(-c_{D}\frac{\dot{\theta}}{\sqrt{2}}+\dot{c}_{-}\right) & = & \left[-i\frac{\kappa}{4}\sin^{2}\theta-\sqrt{g_{0}^{2}+\Omega\left(t\right)^{2}}\right]c_{-}-\left(\frac{i\kappa}{2\sqrt{2}}+\frac{1}{\sqrt{2}}\right)\cos\theta\sin\theta c_{D}\\
 & + & \left(-\frac{i\kappa}{4}\sin^{2}\theta-\frac{1}{2}\cos^{2}\theta+\frac{1}{2}\right)c_{+}
\end{eqnarray*}

Let $c_{0}=c_{+}-c_{-}$, $c_{e}=c_{+}+c_{-}$, we can get

\begin{eqnarray*}
\dot{c}_{D}-\frac{\dot{\theta}}{\sqrt{2}}c_{0} & = & -\frac{\kappa}{2}\cos^{2}\theta c_{D}+\left(\frac{\kappa}{2\sqrt{2}}-\frac{i}{\sqrt{2}}\right)\cos\theta\sin\theta c_{0}\\
\sqrt{2}c_{D}\dot{\theta}+\dot{c}_{0} & = & \left(\frac{\kappa}{\sqrt{2}}-i\sqrt{2}\right)\cos\theta\sin\theta c_{D}-i\sqrt{g_{0}^{2}+\Omega\left(t\right)^{2}}c_{e}+\frac{i}{2}\sin^{2}\theta c_{0}\\
\dot{c}_{e} & = & \left(-\frac{\kappa}{2}\sin^{2}\theta-\frac{i}{2}\sin^{2}\theta\right)c_{e}-i\sqrt{g_{0}^{2}+\Omega\left(t\right)^{2}}c_{0}
\end{eqnarray*}

In the adiabatic approximation, $\dot{\theta}\approx0$, $c_{0}\approx0$
and $c_{e}\approx0$, and $c_{D}\left(0\right)=1$, we can get

\begin{equation}
\dot{c}_{D}=-\frac{\kappa}{2}\cos^{2}\theta c_{D}
\end{equation}

\section{The time-reversible symmetry of the system}

For a closed system, in the basis of $|s\rangle|0\rangle$, $|e\rangle|0\rangle$ and $|g\rangle|1\rangle$,
the wave function of the system is $\psi\left(t\right)=c_{1}\left(t\right)|s\rangle|0\rangle+c_{2}\left(t\right)|e\rangle|0\rangle+c_{3}\left(t\right)|g\rangle|1\rangle$,
applying the systematic Hamiltonian to the schr$\ddot{o}$dinger equation
$i\frac{\partial}{\partial t}|\psi\left(t\right)\rangle=H|\psi\left(t\right)\rangle$,
we can get

\begin{eqnarray}
i\dot{c}_{1}\left(t\right) & = & \Omega\left(t\right)c_{2}\left(t\right)\label{1}\\
i\dot{c}_{2}\left(t\right) & = & \Omega\left(t\right)c_{1}\left(t\right)+g_{0}c_{3}\left(t\right)\\
i\dot{c}_{3}\left(t\right) & = & g_{0}c_{2}\left(t\right)
\end{eqnarray}

Consider the coupling between the cavity mode $a$ and the cavity output or input, $\kappa$ is the cavity decay rate,  the Heisenberg-Langevin function for this open system is

\begin{equation}
\dot{\hat{a}}\left(t\right)=-\frac{i}{\hbar}\left[\hat{a}\left(t\right),H_{I}\right]-\frac{\kappa}{2}\hat{a}\left(t\right)+\sqrt{\kappa}\hat{a}_{in}\left(t\right)
\label{4}
\end{equation}
and the input-output relation is

\begin{equation}
\hat{a}_{in}\left(t\right)+\hat{a}_{out}\left(t\right)=\sqrt{\kappa}\hat{a}\left(t\right)
\label{5}
\end{equation}

We know that, $\left\langle \hat{a}\left(t\right)\right\rangle =c_{3}$.
So that, combine equations $\ref{1}-\ref{5}$, we can get

\begin{eqnarray}
i\dot{c}_{1}\left(t\right) & = & \Omega\left(t\right)c_{2}\left(t\right)\\
i\dot{c}_{2}\left(t\right) & = & \Omega\left(t\right)c_{1}\left(t\right)+g_{0}c_{3}\left(t\right)\\
i\dot{c}_{3}\left(t\right) & = & -\frac{i\kappa}{2}c_{3}\left(t\right)+g_{0}c_{2}\left(t\right)+i\sqrt{\kappa}a_{in}\left(t\right)
\end{eqnarray}

The input-output relation becomes

\begin{equation}
\langle\hat{a}_{in}\left(t\right)\rangle+\langle\hat{a}_{out}\left(t\right)\rangle=\sqrt{\kappa}c_{3}\left(t\right)
\end{equation}

%In the output process, we have $\langle\hat{a}_{in}\left(t\right)\rangle=0$ and $\langle\hat{a}_{out}\left(t\right)\rangle=\sqrt{\kappa}c_{3}\left(t\right)$,
%then there is
%
%\begin{eqnarray*}
%i\dot{c}_{1}\left(t\right) & = & \Omega\left(t\right)c_{2}\left(t\right)\\
%i\dot{c}_{2}\left(t\right) & = & \Omega\left(t\right)c_{1}\left(t\right)+g_{0}c_{3}\left(t\right)\\
%i\dot{c}_{3}\left(t\right) & = & -\frac{i\kappa}{2}c_{3}\left(t\right)+g_{0}c_{2}\left(t\right)
%\end{eqnarray*}

If we inverse the time, which means $t\rightarrow -t$, there is

\begin{eqnarray}
i\dot{c}_{1}\left(-t\right) & = & -\Omega\left(-t\right)c_{2}\left(-t\right)\\
i\dot{c}_{2}\left(-t\right) & = & -\Omega\left(-t\right)c_{1}\left(-t\right)-g_{0}c_{3}\left(-t\right)\\
i\dot{c}_{3}\left(-t\right) & = & \frac{i\kappa}{2}c_{3}\left(-t\right)-g_{0}c_{2}\left(-t\right)-i\sqrt{\kappa}a_{in}\left(-t\right)\label{12}
\end{eqnarray}

and

\begin{equation}
\langle\hat{a}_{in}\left(-t\right)\rangle+\langle\hat{a}_{out}\left(-t\right)\rangle=\sqrt{\kappa}c_{3}\left(-t\right)
\label{13}
\end{equation}

Substitute equation \ref{13} into equation \ref{12}, we can get

\begin{eqnarray}
i\dot{c}_{1}\left(-t\right) & = & -\Omega\left(-t\right)c_{2}\left(-t\right)\\
i\dot{c}_{2}\left(-t\right) & = & -\Omega\left(-t\right)c_{1}\left(-t\right)-g_{0}c_{3}\left(-t\right)\\
i\dot{c}_{3}\left(-t\right) & = & -\frac{i\kappa}{2}c_{3}\left(-t\right)-g_{0}c_{2}\left(-t\right)+i\sqrt{\kappa}a_{out}\left(-t\right)
\end{eqnarray}

If in the input process, $\langle\hat{a}_{in}(t)\rangle$ and $\Omega(t)$ are the time reversal of the output process, the phase difference of $\Omega(t)$ and $g_0$ between the input and output processes are $\frac{\pi}{2}$ and the initial state of the input process equals the final state of the output process, $c_i(t)$ ($i=1,2,3$) will also be the time reversal of the output process. In this model, the phase difference of $\Omega(t)$ and $g_0$ between the input and output processes will not affect the value of $|c_i(t)|$, so that, in our scheme, $\Omega(t)$ and $g_0$ in the input process can be in phase with the output process.

This system has time-reversible symmetry.

\section{absorbing process}

Now, we talk abort the process of absorbing a single photon with pulse shape $f_a(t)$ ($t\in[0,T]$) by using the driving pulse $\Omega_a(t)$, and the probability of absorbing a single photon at time $t=T$ is $1$. 

At time $t=0$, there is $J(T,0)=a_{in}^{\star}\left(0\right)=0$.

Assume that, before a certain time $t_{c}$, there is $J(T,t^{\prime})=a_{in}^{\star}\left(t^{\prime}\right)$.
If at $t_{c}$, there is $J(T,t_c)\neq a_{in}^{\star}\left(t_{c}\right)$,
them we have $c_{D}^{aa}\left(t_{c}\right)\neq\int_{0}^{t_{c}}a_{in}^{\star}\left(t^{\prime}\right)a_{in}\left(t^{\prime}\right)dt^{\prime}$.
From the physical significance, we know that, $c_{D}^{aa}\left(t_{c}\right)$
cannot be larger than $\int_{0}^{t_{c}}a_{in}^{\star}\left(t^{\prime}\right)a_{in}\left(t^{\prime}\right)dt^{\prime}$,
so that there should be $c_{D}^{aa}\left(t_{c}\right)<\int_{0}^{t_{c}}a_{in}^{\star}\left(t^{\prime}\right)a_{in}\left(t^{\prime}\right)dt^{\prime}$,
which means that, at time $t_{c}$, some input pulse is reflected
by the system, and haven't be absorbed by the atom. Then at time $T$,
$c_{D}^{aa}\left(T\right)$ should be smaller than $1$, which violates
the presupposition, so that, we can prove that, at any time $t^{\prime}$,
there should be $J(T,t^{\prime})=a_{in}^{\star}\left(t^{\prime}\right)$.

\nocite{*}

\bibliography{main}

% Full bibliography added automatically for Optics Letters submissions; the following line will simply be ignored if submitting to other journals.
% Note that this extra page will not count against page length

\end{document}